\documentclass[a4paper,11pt]{article}
\usepackage{pos}
\usepackage{physics}

\newcommand{\bt}{\mathbf{b}}
\newcommand{\rt}{\mathbf{r}}

\newcommand{\xpom}{x_\mathbb{P}}

\newcommand{\nc}{N_\mathrm{c}}

\newcommand{\Deltat}{\mathbf{\Delta}}

\title{Probing gluon saturation and nuclear structure\\ in photon-nucleus collisions}
\author*[a,b]{Heikki Mäntysaari}
\affiliation[a]{Department of Physics, University of Jyväskylä, %
 P.O. Box 35, 40014 University of Jyväskylä, Finland}
  \affiliation[b]{Helsinki Institute of Physics, P.O. Box 64, 00014 University of Helsinki, Finland}

\author[c]{Farid Salazar}
\affiliation[c]{Institute for Nuclear Theory, University of Washington, Seattle WA 98195-1550, USA}

\author[d]{Bj\"orn Schenke}
\affiliation[d]{Physics Department, Brookhaven National Laboratory, Upton, NY 11973, USA}

\author[e,f]{Chun Shen}
\affiliation[e]{Department of Physics and Astronomy, Wayne State University, Detroit, Michigan 48201, USA}
\affiliation[f]{RIKEN BNL Research Center, Brookhaven National Laboratory, Upton, NY 11973, USA}

\author[g,h]{Wenbin Zhao}
\affiliation[g]{Nuclear Science Division, Lawrence Berkeley National Laboratory, Berkeley, California 94720, USA}
\affiliation[h]{Physics Department, University of California, Berkeley, California 94720, USA}

\emailAdd{heikki.mantysaari@jyu.fi}
\abstract{
We calculate exclusive vector meson photoproduction within the Color Glass Condensate framework in high-energy photon-nucleus scattering probed experimentally in ultra peripheral heavy ion collisions at RHIC and at the LHC. 
When the free parameters are constrained by the $\gamma+p$ data from HERA, we predict significant nuclear suppression for both the coherent and incoherent photoproduction cross section in the TeV range. Our results indicate that the LHC data prefers even stronger saturation effects at the highest collision energies. Furthermore, we demonstrate how the linear polarization of photons in ulra peripheral collisions generates azimuthal modulations in the decay products of the exclusively produced vector meson. We show how these measurements can probe details of the nuclear geometry, specifically the deformed structure of the uranium nuclei.
}

\FullConference{31st International Workshop on Deep Inelastic Scattering (DIS2024)\\
 8–12 April 2024\\
Grenoble, France\\}

\begin{document}
\maketitle

\section{Introduction}

Ultra peripheral collisions (UPCs) provide access to very high-energy photon-nucleus scattering. In exclusive vector meson photoproduction such as $\gamma + A \to \mathrm{J}/\psi+A$ the photon splitting to quark-antiquark pair can be understood perturbatively, and as such these exclusive processes are both experimentally and theoretically clean. Given that the $\mathrm{J}/\psi$ mass is of the same order as the nuclear saturation scale, $\mathrm{J}/\psi$ photoproduction processes can be expected to be powerful probes of gluon saturation phenomena.

In UPCs there are typically two indistinguishable contributions: it is not known which nucleus emits the photon and which nucleus is the target. Although the nucleus is probed at $\xpom=\frac{M_V}{\sqrt{s}}e^{\pm y}$ where $y$ and $M_V$ refer to the vector meson rapidity and mass, respectively, the high-energy photon flux is heavily suppressed and at forward rapidities the large-$\xpom$ contribution dominates. Recently, however, it has become possible to separate the two contributions following the method developed in Ref.~\cite{Guzey:2013jaa} which relies on measurements of UPC cross sections in different forward neutron multiplicity classes. This has allowed ALICE~\cite{ALICE:2023jgu} and CMS~\cite{CMS:2023snh} to extract the $\gamma+\mathrm{Pb}$ cross sections from the UPC data. 

In this Talk, we summarize our recent work reported in Refs.~\cite{Mantysaari:2022sux,Mantysaari:2023xcu}, where we confront state-of-the-art calculations performed within the Color Glass Condensate framework to describe saturation effects with this high-energy data. Furthermore, we review recent work where instead of separating the $\gamma+\mathrm{Pb}$ cross section one focuses on specific features of the UPC collisions, namely on a genuine interfernce pattern that is visible in the vector meson decay products as a result of the two interfering amplitudes. 

\section{Vector meson photoproduction at high energy}

 \begin{figure*}
       \begin{minipage}{.49\textwidth}
        \centering
        \includegraphics[width=\textwidth]{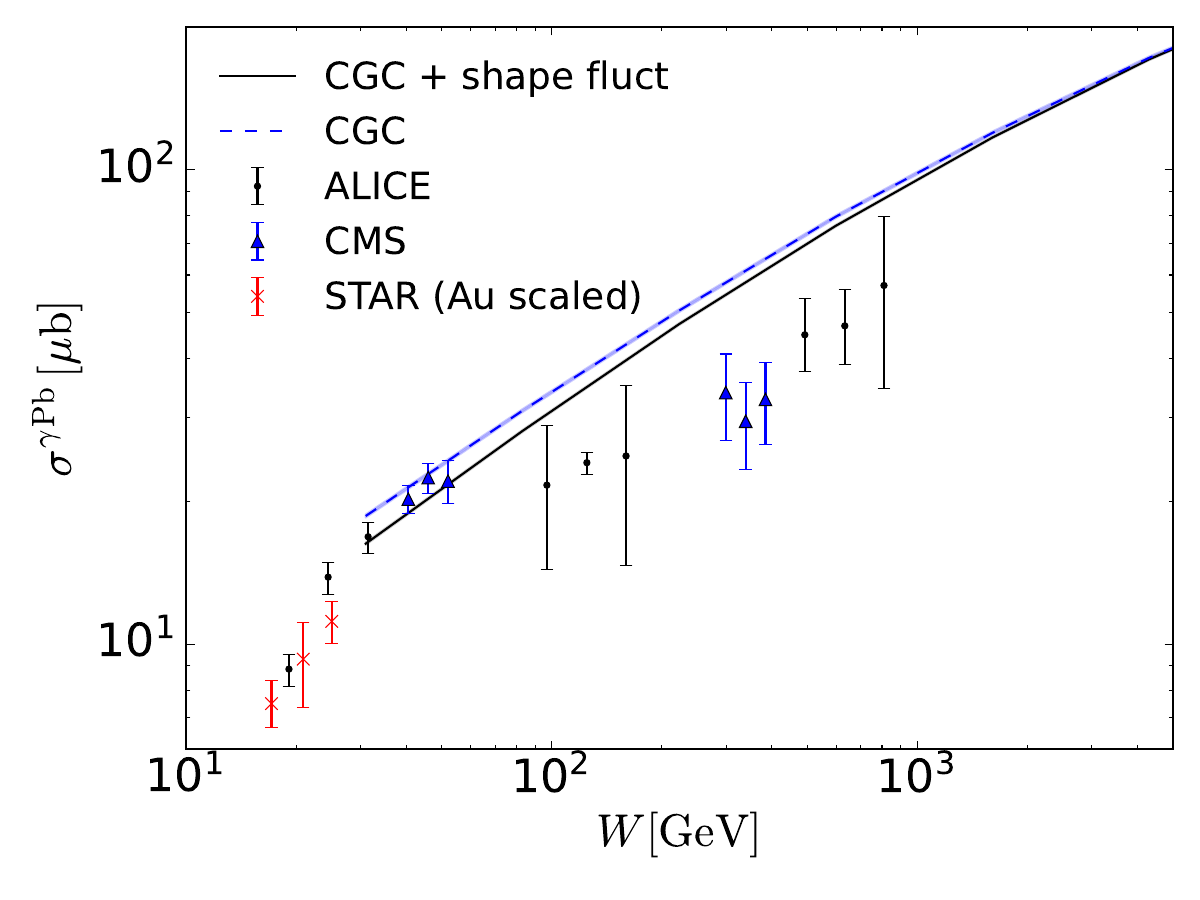}
        \caption{Coherent $\mathrm{J}/\psi$ photoproduction cross section as a function of center-of-mass energy}
        \label{fig:coh_Wdep}
        \end{minipage}
    %\end{figure}
    % \begin{figure}
    \begin{minipage}
        {0.49\textwidth}
       % \begin{figure}
        \centering
        \includegraphics[width=\columnwidth]{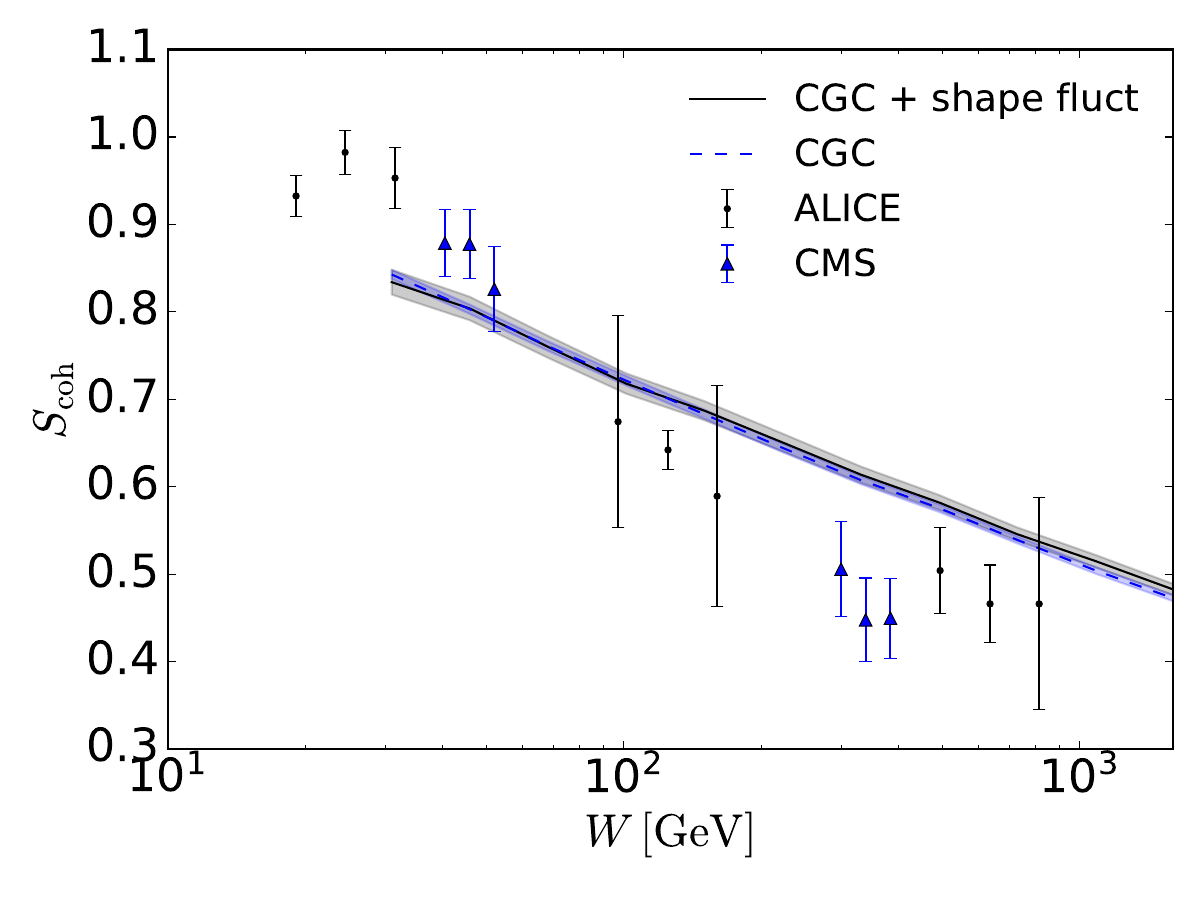}
        \caption{Nuclear modification factor $S_\mathrm{coh}$ for coherent $\mathrm{J}/\psi$ photoproduction as a function energy.}
        \label{fig:Scoh}
        \end{minipage}
    \end{figure*}

    \begin{figure*}
       \begin{minipage}{.49\textwidth}
        \centering
        \includegraphics[width=\textwidth]{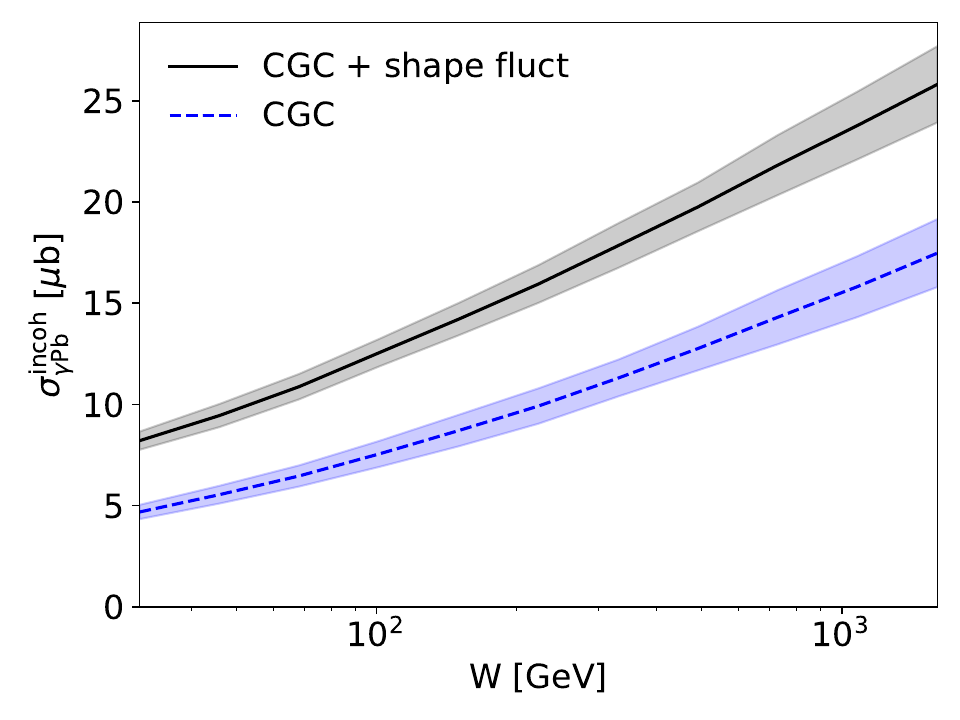}
        \caption{Incoherent $\mathrm{J}/\psi$ photoproduction cross section calculated with and without nucleon substructure fluctuations.}
        \label{fig:inoch_wdep}
        \end{minipage}
    %\end{figure}
    % \begin{figure}
    \begin{minipage}
        {0.49\textwidth}
       % \begin{figure}
        \centering
        \includegraphics[width=\textwidth]{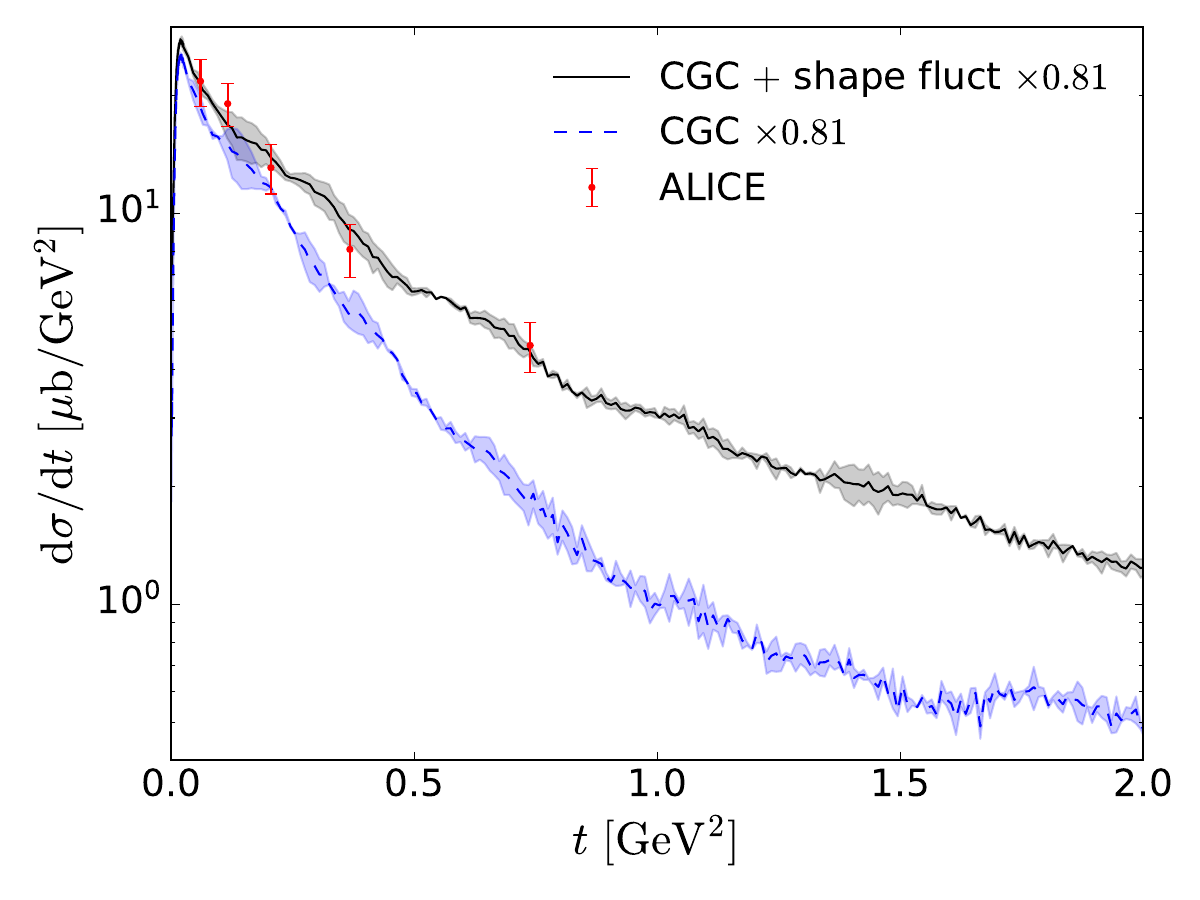}
        \caption{Incoherent $\mathrm{J}/\psi$ photoproduction cross section as a function of squared momentum transfer compared to ALICE data~\cite{ALICE:2023gcs}.}
        \label{fig:incoh_tspectra}
        \end{minipage}
    \end{figure*}

At high energy, it is convenient to describe the scattering process in the dipole picture, where the scattering amplitude can be written as
\begin{equation}
     -i{\mathcal{A}} = \int \dd[2]{\rt} \dd[2]\bt  \int_0^1 \frac{\dd{z}}{4\pi}  [\Psi_V^* \Psi_\gamma](Q^2,\rt,z)
     e^{-i\bt\cdot\Deltat}  N(\rt,\bt,z) \,
\end{equation}
where the dipole-target scattering amplitude is
\begin{equation}
    N(\rt,\bt,z) 
     = 1 - \frac{1}{\nc} \tr \left[ V\left(\bt + (1-z)\rt\right) V^\dagger\left(\bt - z\rt\right) \right]. 
\end{equation}
Here the dipole size and impact parameter are denoted by $\rt$ and $\bt$. The Wilson lines depend implicitly on the momentum fraction $\xpom$ as described by the JIMWLK evolution equation~\cite{Mueller:2001uk}. The coherent cross section where the target remains intact is proportional to the average scattering amplitude: $\frac{\dd \sigma^\mathrm{coh}}{\dd t} = \frac{1}{4\pi} |\langle \mathcal{A}\rangle_\Omega|^2$ where the average is over possible target configurations. Similarly, the incoherent cross section where the target dissociates can be written as $\frac{\dd \sigma^\mathrm{incoh}}{\dd t} = \frac{1}{4\pi} \left[ \langle|\mathcal{A}|^2\rangle_\Omega - |\langle \mathcal{A}\rangle_\Omega|^2 \right]$. As such, the coherent cross section is sensitive to the average structure of the target, and the incoherent cross section probes the event-by-event fluctuations. 

Free parameters of the setup that describe the proton saturation scale,  event-by-event geometry fluctuations geometry, and infrared regulators have been fixed by HERA $\gamma+p\to\mathrm{J}/\psi+p$ data in Ref.~\cite{Mantysaari:2022sux}. There are no free parameters when generalizing the setup from protons to nuclei, except for the standard Woods-Saxon geometry.

The coherent $\gamma+\mathrm{Pb}\to\mathrm{J}/\psi+\mathrm{Pb}$ cross section, compared to the ALICE, CMS and STAR data~\cite{STAR:2023nos}, is shown in Fig.~\ref{fig:coh_Wdep}. The cross section in the small $W<100\,\mathrm{GeV}$ region is well reproduced, but even stronger saturation effects than predicted by our setup would be required to describe the data in the high-energy region. To quantify the magnitude of the saturation effect, we show in Fig.~\ref{fig:Scoh} the nuclear modification factor defined as $
    S_\mathrm{coh} = \sqrt{ \frac{\sigma^{\gamma A} }{ \sigma^\mathrm{IA} }}
$. Here the Impulse Approximation result $\sigma^\mathrm{IA}$ is the result obtained by scaling the $\gamma+p$ cross section to the $\gamma+A$ case by only taking into account the nuclear geometry (form factor). We predict a very strong suppression due to gluon saturation at high energies, but an even stronger suppression would be preferred by the LHC data.

Similarly, we present predictions for the future measurement of the incoherent $\gamma+\mathrm{Pb}\to\mathrm{J}/\psi+\mathrm{Pb}$ cross section as a function of photon-nucleon center-of-mass energy $W$. The results are shown in Fig.~\ref{fig:inoch_wdep}. Here we present predictions obtained with and without nucleon substructure fluctuations (as constrained by HERA $\gamma+p\to\mathrm{J}/\psi+p$ data, see~\cite{Mantysaari:2020axf} for a review). The additional source of fluctuations in the nuclear geometry increases the incoherent cross section at large $|t|$ and consequently also the $t$ integrated incoherent cross section. This is illustrated in Fig.~\ref{fig:incoh_tspectra}, where the incoherent cross section as a function of squared momentum transfer is shown. Finally, the nuclear modification factor for the incoherent cross section (the ratio to the impulse approximation without a square root) is shown in Fig.~\ref{fig:incoh_ia_ratio}. This ratio is defined as in Ref.~\cite{Kryshen:2023bxy}, and again a significant suppression at high energies is predicted. 

\section{Interference patterns}

\begin{figure*}
       \begin{minipage}{.49\textwidth}
        \centering
        \includegraphics[width=\columnwidth]{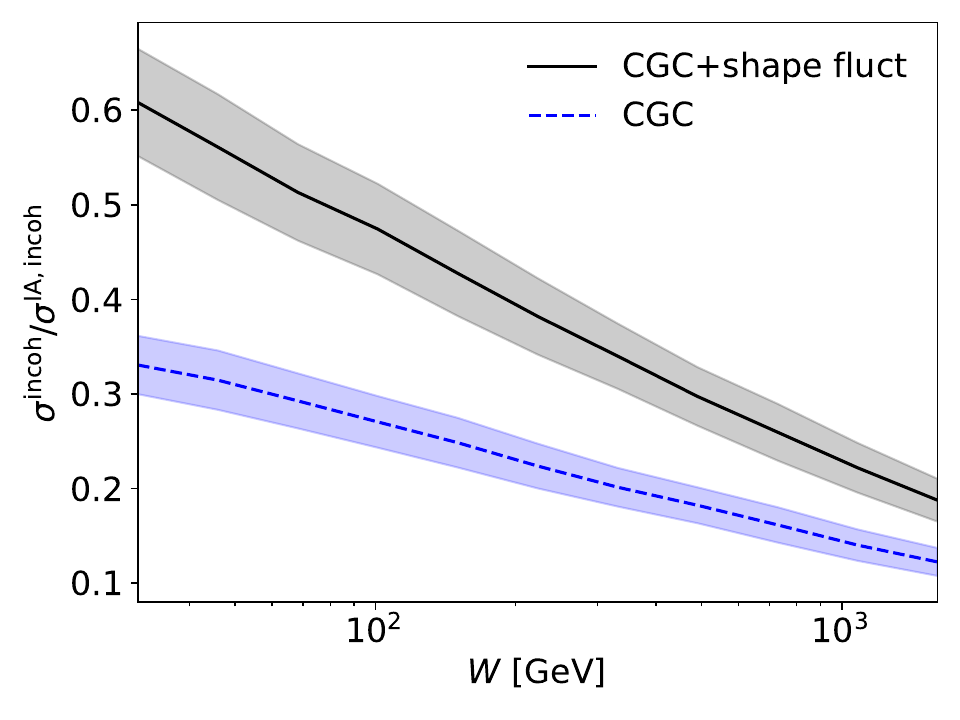}
        
        \caption{Incoherent-to-coherent $\mathrm{J}/\psi$ photoproduction cross section ratio in $\gamma+\mathrm{Pb}$ scattering.}
        \label{fig:incoh_ia_ratio}
        \end{minipage}
    %\end{figure}
    % \begin{figure}
    \begin{minipage}
        {0.49\textwidth}
       % \begin{figure}
        \centering
        \includegraphics[width=\columnwidth]{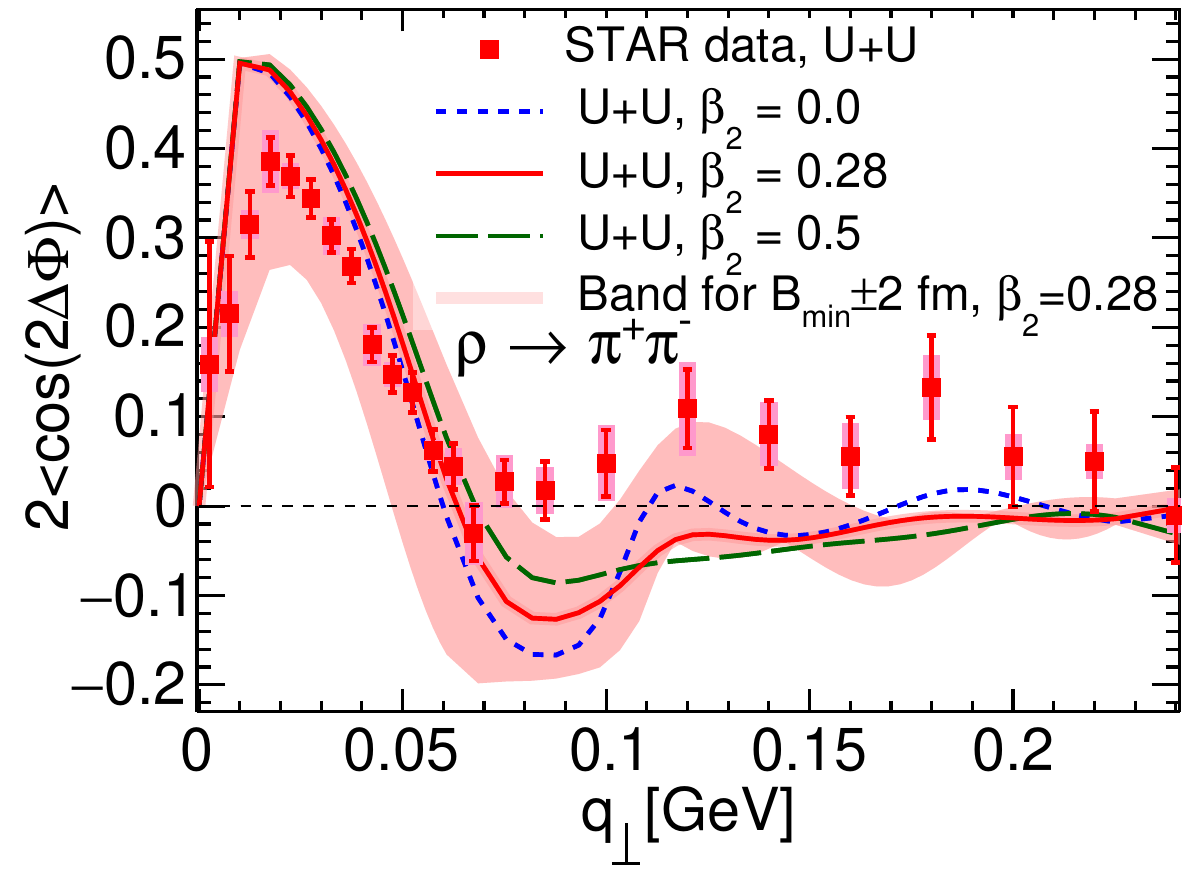}
        \caption{Azimuthal modulation of the $\rho$ decay products in $\mathrm{U}+\mathrm{U}\to \mathrm{U}+\mathrm{U} + (\rho \to \pi^+ \pi^-)$ as a function of $\rho$ transverse momentum $q_\perp$.}
        \label{fig:v2}
        \end{minipage}
    \end{figure*}

The two interfering amplitudes and the fact that the photons are linearly polarized generate a non-trivial interference pattern in the decay products of the exclusively produced vector meson. In Ref.~\cite{Mantysaari:2023prg} we calculate these angular correlations using approximatively the same Color Glass Condensate based setup as the one used above in studies of gluon saturation phenomena. In Fig.~\ref{fig:v2} we show the calculated $2\langle \cos(2\Delta\Phi)\rangle$ modulation coefficient, where $\Delta\Phi$ is the angle between the transverse momentum of the exclusively produced $\rho$ meson, $q_\perp$, and one of its decay product pions.  The result shown in Fig.~\ref{fig:v2} for the Uranium-Uranium collisions demonstrates that this modulation is sensitive to the details of the nuclear geometry, specifically to the deformation parameter $\beta_2$. As discussed in detail in Ref.~\cite{Mantysaari:2023prg}, this sensitivity in deformations is mostly due to the fact that the minimum impact parameter between the nuclei required for an UPC event depends on deformations.

\section{Conclusions}

Very high-energy photon-nucleus collisions measured in ultra peripheral collisions at the LHC show signs of very strong saturation effects. These effects are even stronger than what is naturally predicted by the applied (leading order) CGC calculations. Future measurements of the energy-dependent incoherent cross section will provide further constraints to the applied models. In addition to vector meson production cross sections, we have also highlighted the possibilities arising from more differential studies of UPC scattering processes, in this case focusing on azimuthal modulations in the decay products of the exclusively produced vector meson demonstrating that such correlations are sensitive probes of the nuclear geometry.

\begin{acknowledgments}
H.M. is supported by the Research Council of Finland, the Centre of Excellence in Quark Matter, and projects 338263 and 346567, and under the European Union’s Horizon 2020 research and innovation programme by the European Research Council (ERC, grant agreements No. ERC-2023-101123801 GlueSatLight and No. ERC-2018-ADG-835105 YoctoLHC) and by the STRONG-2020 project (grant agreement No. 824093). B.P.S. is supported by the U.S. Department of Energy, Office of Science, Office of Nuclear Physics, under DOE Contract No.~DE-SC0012704  and within the framework of the Saturated Glue (SURGE) Topical Theory Collaboration. 
F.S. is supported by the Institute
for Nuclear Theory’s U.S. DOE under Grant No. DE-FG02-00ER41132.
%Computing resources from CSC – IT Center for Science in Espoo, Finland and the Finnish Grid and Cloud Infrastructure (persistent identifier \texttt{urn:nbn:fi:research-infras-2016072533}) were used in this work.
The content of this article does not reflect the official opinion of the European Union and responsibility for the information and views expressed therein lies entirely with the authors. 
\end{acknowledgments}

\bibliographystyle{JHEP-2mod.bst}
\bibliography{refs}

\end{document}